\documentclass{article}
\usepackage{subcaption, multirow, multicol, booktabs,amsfonts, kotex}  
\usepackage{spconf,amsmath,graphicx}
\usepackage[table]{xcolor}



\title{STaR: Distilling Speech Temporal Relation for\\Lightweight Speech Self-Supervised Learning Models} 
\name{Kangwook Jang$^1$, Sungnyun Kim$^2$, Hoirin Kim$^1$}
\address{
  $^1$School of Electrical Engineering, KAIST \\
  $^2$Kim Jaechul Graduate School of AI, KAIST}
\begin{document}
\ninept    
\maketitle
%

\begin{abstract}

Albeit great performance of Transformer-based speech self-supervised learning\,(SSL) models, their large parameter size and computational cost make them unfavorable to utilize.
In this study, we propose to compress the speech SSL models by distilling \textit{speech temporal relation}\,(STaR).
Unlike previous works that directly match the representation for each speech frame, STaR distillation transfers temporal relation between speech frames, which is more suitable for lightweight student with limited capacity.
We explore three STaR distillation objectives and select the best combination as the final STaR loss.
Our model distilled from HuBERT\,\textsc{Base} achieves an overall score of 79.8 on SUPERB benchmark, the best performance among models with up to 27 million parameters.
We show that our method is applicable across different speech SSL models and maintains robust performance with further reduced parameters.
\end{abstract}


\setlength{\footnotesep}{0pt}
\addtolength{\skip\footins}{-0.3pt}
\let\thefootnote\relax\footnotetext{This work was supported by the National Research Foundation of Korea grant funded by the Korea government\,(MSIT)\,(No. 2021R1A2C1014044).}

\section{Introduction}
\label{sec:intro}

Transformer-based speech self-supervised learning (SSL) models \cite{baevski2020wav2vec, hsu2021hubert, chen2021wavlm} have risen to prominence with great performance in various speech-related tasks\,\cite{yang21c_interspeech, feng2023superb}.
Nonetheless, the downside of these models mainly comes from the requirement of substantial computational resources during the pre-training stage---HuBERT\,\textsc{Base} takes over 82 GPU-days for pre-training, and 32 GPUs en masse are utilized to shorten this\,\cite{hsu2021hubert}.
Another downside is their large parameter size which makes on-device application difficult and renders the speech SSL models unfavorable for many practical scenarios.
The above issues make compression techniques essential for the speech SSL models, and several studies have attempted to solve these issues by applying pruning\,\cite{lai2021parp} or quantization\,\cite{wang2022deep} technique.

Alternatively, knowledge distillation is another approach for model compression that trains a \textit{student} model with a smaller parameter size to imitate the behavior of a larger \textit{teacher} model by matching the student's representation to the teacher's.
Knowledge distillation of the Transformer-based speech SSL models is being actively studied, with previous works of task-specific compression including automatic speech recognition\,(ASR)\,\cite{choi2023masked}, keyword spotting (KWS)\,\cite{lim2023lightweight}, and automatic speaker verification\,(ASV)\,\cite{heo23_interspeech}.
However, given that the speech SSL models are utilized in various downstream tasks, their approaches are limited to only a specific task.

On the other hand, approaches that realize task-agnostic compression via knowledge distillation include DistilHuBERT\,\cite{chang2021distilhubert} and FitHuBERT\,\cite{lee2022fithubert}, where the former suggests shallow and wide student model design, and the latter suggests deep and narrow one.
ARMHuBERT\,\cite{jang23_interspeech} proposes to reuse attention maps across Transformer\,\cite{vaswani2017attention} layers and utilizes both masked and unmasked speech frames for masking distillation.
Some approaches\,\cite{peng23c_interspeech, wang23da_interspeech} jointly conduct distillation with $L_0$ regularization\,\cite{louizos2018learning} and structured pruning.
LightHuBERT\,\cite{wang2022lighthubert} implements masking distillation and architecture search, but it stands apart from other methods due to its excessive computational demand for creating a teacher-sized supernet.

While the aforementioned studies have demonstrated promising results, two limitations still exist.
First, most approaches neglect the weak representation capacity of the student and directly match the complex teacher's representation for each speech frame by introducing additional linear heads\,\cite{chang2021distilhubert, lee2022fithubert, jang23_interspeech, peng23c_interspeech, ashihara2022deep}.
This can be an over-constraint for the lightweight student model, necessitating the establishment of an alternative distillation objective that better suits to the student.
Some studies even inefficiently discard these trained linear heads after distillation, although they can convey the teacher's knowledge\,\cite{chang2021distilhubert, lee2022fithubert, peng23c_interspeech, ashihara2022deep}.
Second, while pruning allows us to manage the size of model parameters by a sparsity ratio, it cannot determine computational cost of the model at the same time.
As a result, computational overhead can be higher than the vanilla distillation approaches, in which the model is specified before the training.

In this study, we explore effective distillation objectives that capture temporal relation between speech frames for lightweight student model.
Additional parameters are not necessary during distillation, enabling the construction of a more compact and computationally efficient student model.
We verify the task-agnostic compression of our proposed objectives using SUPERB benchmark\,\cite{yang21c_interspeech}.
Our model distilled from HuBERT\,\textsc{Base} achieves the best overall score\,\cite{chen2021wavlm} of 79.8 among models with $\sim$\,27 million parameters, while requiring only 30.7\% multiply-accumulates\,(MACs) and 28.1\% parameters of its teacher model.
It even surpasses LightHuBERT\,\cite{wang2022lighthubert}, which demands extensive computational resources for compression, in terms of overall score, number of parameters, and MACs.

\section{Speech Temporal Relation Distillation}
\label{sec:temporal-pattern-distillation}

\begin{figure*}[t]
    \centering
    \includegraphics[width=\textwidth]{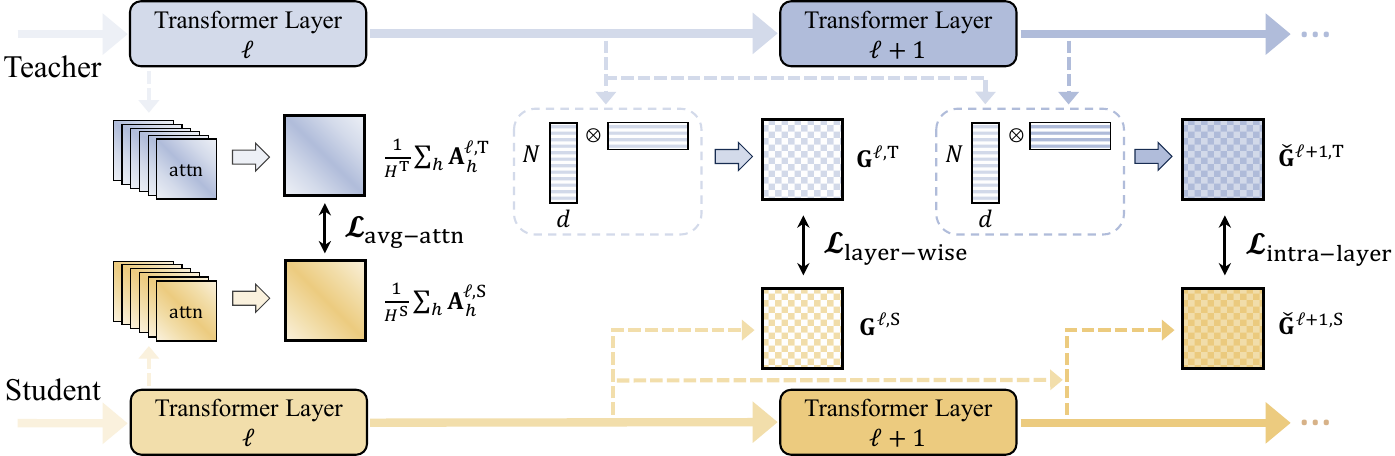}
    \caption{
    We propose three STaR distillation objectives: average attention map, layer-wise TGM, and intra-layer TGM.
    TGM captures the temporal relation by aggregating the channel information at two time steps.
    Each individual loss is summed up across all Transformer layers.
            }
    \label{fig:overview}
    \vspace{-4.5pt}
\end{figure*}

Speech SSL models are capable of representing speech frames into features that function as specific acoustic units for every time step.
This can be attributed to the pre-training scheme of a speech SSL model, predicting the corresponding cluster of each masked frame\,\cite{hsu2021hubert, chen2021wavlm}.
The pre-processing procedure of wav2vec-U\,\cite{baevski2021unsupervised} also involves grouping the speech frames from the same cluster for phonemic segmentation.
As such, speech SSL models generate the representations for each frame that are closely tied to specific acoustic units, and how to represent these units is the core knowledge of the models.
However, directly learning the teacher's representations, which convey the characteristics of these units, can be an over-constraint for the student with limited representation capacity.

To this end, we propose to distill the knowledge from the teacher in a more flexible manner by focusing on the \textit{relation} between speech frames instead.
We explore three \textbf{\underline{S}peech \underline{T}empor\underline{a}l \underline{R}elation (STaR)} distillation objectives (Fig.\,\ref{fig:overview}), by which the distillation transfers pairwise temporal relation of the speech frames.
We first suggest average attention map distillation, a naïve way to distill the temporal relation.
Then, temporal Gram matrix\,(TGM) is introduced as a distillation target, and on top of this, we propose two TGM distillation objectives: layer-wise TGM distillation and intra-layer TGM distillation.
We highlight that our methods are task-agnostic compression and do not require additional parameters during distillation stage unlike previous works\,\cite{chang2021distilhubert, lee2022fithubert, jang23_interspeech, peng23c_interspeech, ashihara2022deep}.

\subsection{Average Attention Map Distillation}
\label{sec:attention-map-distillation}

Attention map in Transformer\,\cite{vaswani2017attention} structure captures the temporal relation between key and query.
Each entry in the attention map indicates a level of the relationship between two associated frames of the sequence, therefore, distilling this map can be an intuitive way to transfer the temporal relation from the teacher.
Given key and query matrices for head $h$, $\textbf{K}_h$, $\textbf{Q}_h \in \mathbb{R}^{d_h \times N}$, in multi-head self-attention\,(MHSA), the corresponding attention map $\textbf{A}_h \in \mathbb{R}^{N \times N}$ is $\textbf{A}_h = \textit{softmax}\big(\textbf{Q}^\top_h\textbf{K}_h / \sqrt{d_h}\big)$.
$d_h$ is the width of the key matrix for head $h$, and $N$ is the sequence length.
Several studies have leveraged the attention maps of all heads in the last Transformer layer for knowledge distillation\,\cite{wang2020minilm, fang2021compressing};
nevertheless, this is not ideal for the speech SSL models where each layer plays an essential role in different speech-related task\,\cite{chen2021wavlm, yang21c_interspeech}. 
Distilling the attention maps of all heads across all layers would be an ideal way, but it leads to excessive computational overhead for training.

Accordingly, we propose to distill the attention map averaged across all heads from each Transformer layer as an alternative.
From each Transformer layer, the proposed loss calculates Kullback-Leibler\,(KL) divergence between the average attention maps of teacher T and student S.
\begin{align}\label{eq:attention-map-distillation}
    \mathcal{L}_{\text{avg-attn}} &= \sum_{\ell=1}^{L}\sum_{t=1}^{N}D_{KL}
        \left(\frac{1}{H^{\text{T}}}\sum_{h=1}^{H^{\text{T}}}\textbf{A}^{\ell, \text{T}}_{h, t}
        \bigg\lVert
        \frac{1}{H^\text{S}} \sum_{h=1}^{H^\text{S}} \textbf{A}^{\ell, \text{S}}_{h, t}\right),
\end{align}
where $H$ and $L$ denote the total number of attention heads and Transformer layers, respectively.
$\textbf{A}^{\ell}_{h, t}$ represents the attention distribution at time step $t$ within the attention head $h$ of Transformer layer $\ell$.

\subsection{Temporal Gram Matrix Distillation}

While distilling the attention maps can transfer their temporal relation to the student, it might not provide sufficient hints since these maps are not the direct features used for inference.
In order to provide stronger hints, we present the distillation objectives regarding temporal relation within each Transformer\,\cite{vaswani2017attention} layer output by computing a Gram matrix.
Correlation between two samples can be expressed via Gram matrix, defined as an inner product between sample representations\,\cite{gatys2015texture}.
Let $\textbf{F} \in \mathbb{R}^{d \times N}$ be a representation of the channel width $d$.
Then, the Gram matrix $\Tilde{\textbf{G}}$ is
\begin{align}\label{eq:original-gram-matrix}
    \Tilde{G}_{ij} = \sum_{k}F_{ik}F_{jk}, \quad \tilde{\textbf{G}} \in \mathbb{R}^{d \times d},
\end{align}
where $F_{\cdot k}$ denotes the $k$-th time step's representation.

Gram matrix has been utilized in speech and vision domain to represent the properties of data, referred to as style.
Beginning with the proposal of artistic style transfer via Gram matrix\,\cite{gatys2015texture}, numerous studies advocate leveraging the matrix for speaker embedding\,\cite{stafylakis21_interspeech, saito2021perceptual} or voice conversion\,\cite{li2020non, shi2022u}.
The style loss\,\cite{gatys2015texture} minimizes the correlation between channels by aggregating positional information within a sample, as in Eq.\,\ref{eq:original-gram-matrix}.
The key strength of this loss lies in its objective which can be seen as a distribution alignment process, minimizing maximum mean discrepancy\,\cite{gretton2012kernel} between two sets\,\cite{li2017demystifying}.

Gram matrix or its variant is also exploited as an objective for conducting knowledge distillation\,\cite{yim2017gift, zhou2022efficient}.
Prior works share the idea that they aggregate temporal or spatial information to illustrate the relation between channels when computing the Gram matrix.
In contrast, we propose a \textbf{temporal Gram matrix\,(TGM)} which aggregates channel information at two time steps to take into account temporal relation between speech frames.
Note that this differs from other works\,\cite{peng2019correlation, jin2023cross} which capture relation between samples within the same batch.
The temporal Gram matrix $\textbf{G}$ is defined as
\begin{align}\label{eq:temporal-gram-matrix}
    {G}_{ij} = \sum_{k}F_{ki}F_{kj}, \quad \textbf{G} \in \mathbb{R}^{N \times N}.
\end{align}

\noindent\textbf{Layer-wise TGM distillation}\quad Layer-wise TGM distillation takes the TGM of each Transformer layer output as an objective.
We also include the first Transformer layer input to distill the front-end convolutional layers more directly.
This approach effectively transfers the information encoded in each layer of the speech SSL models\,\cite{pasad2023comparative} to the corresponding layer of the student model.
Indeed, previous studies have confirmed the effectiveness of layer-wise distillation for the speech SSL models that targets intermediate layers\,\cite{lee2022fithubert, jang23_interspeech, ashihara2022deep}.
Regarding the first Transformer layer input as the zeroth output, the layer-wise TGM distillation loss is the mean squared error\,(MSE) between the TGMs of teacher and student across all layers.
\begin{align}\label{eq:layer-wise-tgm-distillation}
    \mathcal{L}_{\text{layer-wise}} &= \sum_{\ell=0}^{L} \big\lVert \textbf{G}^{\ell, \text{T}}-\textbf{G}^{\ell, \text{S}} \big\rVert^2_2
\end{align}

\noindent\textbf{Intra-layer TGM distillation}\quad To offer a more flexible distillation objective for the student model with reduced parameter size, we propose intra-layer TGM distillation.
Inspired by a flow of solution procedure matrix\,\cite{yim2017gift}, we modify the TGM as computing the temporal relation between the input and output of a single Transformer layer, specifically emphasizing its intra-layer role.
This matrix captures the progression of speech representation within every individual layer, providing a more flexible objective based on two different representations.
The modified TGM $\check{\textbf{G}}$ of the Transformer layer $\ell$ and the intra-layer TGM distillation loss are defined as
\begin{align}
    \label{eq:redefined-tgm}
    \check{G}^{\ell}_{ij} &= \sum_{k}F^{\ell-1}_{ki}F^{\ell}_{kj}, \quad \check{\textbf{G}}^{\ell} \in \mathbb{R}^{N \times N}, \\
    \label{eq:intra-layer-tgm-distillation}
    \mathcal{L}_{\text{intra-layer}} &= \sum_{\ell=1}^{L} \big\lVert \check{\textbf{G}}^{\ell, \text{T}}-\check{\textbf{G}}^{\ell, \text{S}} \big\rVert^2_2.
\end{align}

The advantage of the objectives introduced in this section is needlessness of additional parameters for implementing distillation.
Even if the channel width of the student model is not equal to the teacher's, our losses can be formulated without additional linear projections, provided their time lengths $N$ are the same.
As a result, we can create a more compact student and fully transfer the teacher's temporal relation knowledge, unlike previous works that discard the projection heads carrying the teacher's knowledge\,\cite{chang2021distilhubert, lee2022fithubert, peng23c_interspeech, ashihara2022deep}.

\section{Results}
\label{sec:page}

\subsection{Experimental Details}

\textbf{Training details}\quad
HuBERT\,\textsc{Base}\,\cite{hsu2021hubert}, comprising 12 Transformer \cite{vaswani2017attention} layers, is our primary speech SSL model to distill.
We train our student model with LibriSpeech\,\cite{panayotov2015librispeech} train-clean-100 dataset\,(100h) for 200 epochs, and full dataset\,(960h) for 100 epochs.
We follow the student's training recipe of \cite{jang23_interspeech}, except for the cosine scheduler and the learning rate of 1e-3.
Effective batch size including gradient accumulation is 48, using two NVIDIA RTX 4090 GPUs.

\noindent\textbf{Student description}\quad
We create the student model by reducing the width of attention layer and feed-forward network\,(FFN) while retaining the number of layers in the teacher model.
This is to perform layer-to-layer\,(L2L) distillation for every layer, as in Eqs.\,\ref{eq:attention-map-distillation},\,\ref{eq:layer-wise-tgm-distillation}, and\,\ref{eq:intra-layer-tgm-distillation}.
The width of (attention, FFN) is set as (432, 976) or (432, 1392) to match the parameter size with prior works\,\cite{jang23_interspeech, peng23c_interspeech, wang2022lighthubert}.

\noindent\textbf{Evaluation}\quad
Once distilled, the student model is evaluated on the SUPERB benchmark\,\cite{yang21c_interspeech} to verify its task-agnostic characteristic.
The benchmark consists of 10 speech-related tasks, including phoneme recognition\,(PR), ASR, KWS, query by example spoken term detection\,(QbE), speaker identification\,(SID), ASV, speaker diarization\,(SD), intent classification\,(IC), slot filling\,(SF), and emotion recognition\,(ER).
We fine-tune the model with default SUPERB recipes, except for enabling the learning rate scheduler and setting the learning rate of the SID task as 5e-3.
For the metrics that assess the entire tasks, we use overall score from\,\cite{chen2021wavlm} and generalizability score from\,\cite{feng2023superb}; the former is the absolute average of each task's performance, while the latter uses the linearly mapped scores in the range 0\,(log mel filterbank)\,$\sim$\,1000\,(SOTA reported in \cite{yang21c_interspeech}).
We follow the implementation of \cite{feng2023superb} for estimating MACs.

\subsection{Selection of STaR Loss}

Table\,\ref{tab:exploring-star-table} compares the performance of the student models including our proposed STaR distillations.
Average attention map distillation outperforms FitHuBERT\,\cite{lee2022fithubert} and is on par with MaskHuBERT\,\cite{jang23_interspeech}, while requiring 4.6\% and 16.3\% fewer parameters than FitHuBERT and MaskHuBERT, respectively.
It also reveals that L2L distillation for every layer is more effective than addressing only the last layer\,($\mathcal{L}_\text{last-attn}$)\,\cite{wang2020minilm}.
Besides, layer-wise and intra-layer TGM distillation outperform the average attention map distillation, as matching the TGMs synchronizes temporal relation of the features that serve as direct outputs in fine-tuning.
Incorporating the consideration of the average rank\,\cite{chang2021distilhubert} among the STaR students, we set the default STaR loss as the combination of layer-wise and intra-layer losses and denote this student as STaRHuBERT.
\begin{table}[!t]
    \centering
    \vspace{-6pt}
    \caption{
    Results of the SUPERB benchmark for different STaR distillation objectives.
    ``Overall'' and ``General.'' denote the overall and generalizability scores.
    ``Ranking'' is the average rank over all SUPERB tasks proposed in \cite{chang2021distilhubert}.
    $\mathcal{L}_{\text{last-attn}}$ is the last layer's attention map distillation\,\cite{wang2020minilm}.
    We have reproduced MaskHuBERT-100h\,\cite{jang23_interspeech}.
    }
    \label{tab:exploring-star-table}
    \vspace{4.5pt}
    \addtolength{\tabcolsep}{1pt}
    \resizebox{\linewidth}{!}{
    \begin{tabular}{lccc}
    \toprule
    Methods & Overall\,$\uparrow$ & General.\,$\uparrow$ & Ranking\,$\downarrow$  \\
    \midrule
    \textit{\textcolor{gray}{LibriSpeech 100h distillation}} \\
    FitHuBERT\,\cite{lee2022fithubert} & 74.5 & 695 & - \\
    MaskHuBERT & 76.3 & 789 & - \\
    $\mathcal{L}_{\text{last-attn}}$ & 75.7 & 750 & - \\
    $\mathcal{L}_{\text{avg-attn}}$ & 76.5 & 766 & 4.4 \\
    $\mathcal{L}_{\text{layer-wise}}$ & \textbf{77.8} & 829 & 2.7 \\
    $\mathcal{L}_{\text{intra-layer}}$  & 77.7 & 820 & 3.2 \\
    $\mathcal{L}_{\text{layer-wise}} + \mathcal{L}_{\text{intra-layer}}$  & \textbf{77.8} & \textbf{831} & \textbf{2.2} \\
    $\mathcal{L}_{\text{layer-wise}} + \mathcal{L}_{\text{intra-layer}} + \mathcal{L}_{\text{avg-attn}}$ & 77.7 & 827 & 2.6 \\
    \midrule
    \textit{\textcolor{gray}{LibriSpeech 960h distillation}} \\
    $\mathcal{L}_{\text{layer-wise}}$  & 79.4 & \textbf{887} & 2.1 \\
    $\mathcal{L}_{\text{layer-wise}} + \mathcal{L}_{\text{intra-layer}}$ & \textbf{79.5} & \textbf{887} & \textbf{1.7} \\
    $\mathcal{L}_{\text{layer-wise}} + \mathcal{L}_{\text{intra-layer}} + \mathcal{L}_{\text{avg-attn}}$ & 78.8 & 871 & 2.3\\
    \bottomrule
    \end{tabular}
    }
    \vspace{-4.5pt}

\end{table}
\begin{table*}[!t]
    \centering
    \caption{
    Detailed evaluation results on the SUPERB benchmark. 
    Metrics include number of parameters, number of MACs, phoneme error rate\,(PER,\%), word error rate\,(WER,\%) without language model, accuracy\,(Acc,\%), maximum term weighted value\,(MTWV), equal error rate\,(EER,\%), diarization error rate\,(DER,\%), F1 score\,(F1,\%), and concept error rate\,(CER,\%).
    The best performance is bolded, and the second place is underlined.
    LightHuBERT\,\cite{wang2022lighthubert} stands apart from other works, requiring extensive computational resources for compression.}
    \label{tab:full-superb-benchmark}
    \vspace{4.5pt}
    \addtolength{\tabcolsep}{-1pt}
    \resizebox{\textwidth}{!}{
    \begin{tabular}{lccccccccccccccc}
    \toprule
    & \multicolumn{2}{c}{Computation} & \multicolumn{2}{c}{Performance} & \multicolumn{4}{c}{Content} &  \multicolumn{3}{c}{Speaker} & \multicolumn{3}{c}{Semantics} & Paral. \\
    \cmidrule(l{2pt}r{2pt}){2-3} \cmidrule(l{2pt}r{2pt}){4-5} \cmidrule(l{2pt}r{2pt}){6-9} \cmidrule(l{2pt}r{2pt}){10-12} \cmidrule(l{2pt}r{2pt}){13-15} \cmidrule(l{2pt}r{2pt}){16-16}
    & Params & MACs & Overall & General. & PR & ASR & KS & QbE & SID & ASV & SD & IC & \multicolumn{2}{c}{SF} & ER \\
    \cmidrule(l{2pt}r{2pt}){2-2} \cmidrule(l{2pt}r{2pt}){3-3} \cmidrule(l{2pt}r{2pt}){4-4} \cmidrule(l{2pt}r{2pt}){5-5} \cmidrule(l{2pt}r{2pt}){6-6} \cmidrule(l{2pt}r{2pt}){7-7} \cmidrule(l{2pt}r{2pt}){8-8} \cmidrule(l{2pt}r{2pt}){9-9} \cmidrule(l{2pt}r{2pt}){10-10} \cmidrule(l{2pt}r{2pt}){11-11} \cmidrule(l{2pt}r{2pt}){12-12} \cmidrule(l{2pt}r{2pt}){13-13} \cmidrule(l{2pt}r{2pt}){14-15} \cmidrule(l{2pt}r{2pt}){16-16}
    Models & \negthickspace Millions\,$\downarrow$\negthickspace & Giga\,$\downarrow$ & Score\,$\uparrow$ & Score\,$\uparrow$ & PER\,$\downarrow$ & WER\,$\downarrow$ & Acc\,$\uparrow$ & MTWV\,$\uparrow$ & Acc\,$\uparrow$ & EER\,$\downarrow$ & DER\,$\downarrow$ & Acc\,$\uparrow$ & F1\,$\uparrow$ & CER\,$\downarrow$ & Acc\,$\uparrow$ \\
    \midrule
    \textit{\textcolor{gray}{Baselines}}   \\
    SOTA\,\cite{yang21c_interspeech} & - & - & 82.8 & 1000 & 3.53 & 3.62 & 96.66 & 0.0736 & 90.33 & 5.11 & 5.62 & 98.76 & 89.81 & 21.76 & 67.62 \\
    FBANK\,(log mel filterbank) & 0 & 0.4791 & 46.5 & 0 & 82.01 & 23.18 & 8.63 & 0.0058 & 8.5E-4 & 9.56 & 10.55 & 9.1 & 69.64 & 52.94 & 35.39 \\
    HuBERT\,\textsc{Base} \cite{hsu2021hubert} & 94.70 & 1669 & 80.8 & 941 & 5.41 & 6.42 & 96.30 & 0.0736 & 81.42 & 5.11 & 5.88 & 98.34 & 88.53 & 25.20 & 64.92 \\
    LightHuBERT\,$a_\text{Small}$\,\cite{wang2022lighthubert} & 27.00 & 860.7 & 79.1  & 900 & 6.60 & 8.33 & 96.07 & 0.0764 & 69.70 & 5.42 & 5.85 & 98.23 & 87.58 & 26.90 & 64.12 \\
    \midrule
    \multicolumn{5}{l}{\textit{\textcolor{gray}{LibriSpeech 960h distillation}}} & \\
    ARMHuBERT-S\,\cite{jang23_interspeech} & 22.39 & 449.4 & 77.5 & 828 & 8.63 & 10.82 & 96.82 & 0.0720 & 63.76 & 5.58 & 7.01 & 97.02 & 86.34 & 29.02 & 62.96 \\
    DPHuBERT\,\cite{peng23c_interspeech} & 23.59 & 654.1 & 78.9 & 866 & 9.67 & 10.47 & 96.36 & 0.0693 & 76.83 & 5.84 & 5.92 & 97.92 & 86.86 & 28.26 & 63.16 \\
    \textbf{STaRHuBERT\,(ours)} & \textbf{22.31} & \underline{463.5} & \underline{79.5} & \underline{887} & \underline{8.16} & \underline{9.35} & 96.27 & 0.0688 & \underline{77.58} & \textbf{5.39} & 6.05 & \underline{97.55} & \underline{87.94} & \textbf{25.31} & 63.01 \\
    \textbf{STaRHuBERT-L\,(ours)} & 26.63 & 511.9 & \textbf{79.8} & \textbf{896} & \textbf{7.97} & \textbf{8.91} & \underline{96.56} & 0.0677 & \textbf{78.66} & \underline{5.45} & \textbf{5.83} & 97.50 & \textbf{88.01} & \underline{25.36} & \textbf{63.48}  \\
    \midrule
    \multicolumn{5}{l}{\textit{\textcolor{gray}{LibriSpeech 100h distillation}}} & \\
    ARMHuBERT-S\,\cite{jang23_interspeech} & 22.39 & 449.4 & 76.8 & 800 & 9.17 & 11.83 & 96.01 & 0.0569 & 66.48 & 5.92 & 6.23 & 95.97 & 83.89 & 33.29 & 63.29 \\
    DPHuBERT\,\cite{peng23c_interspeech} & 23.59 & 654.1 & 77.7 & 819 & 10.02 & 11.38 & 96.36 & 0.0634 & 73.37 & 6.25 & 6.03 & 97.42 & 84.83 & 33.03 & 62.78 \\
    \textbf{STaRHuBERT\,(ours)} & \textbf{22.31} & 463.5 & \textbf{77.8} & \textbf{831} & \textbf{9.17} & \textbf{10.92} & \textbf{96.46} & 0.0626 & 72.26 & \textbf{5.66} & \textbf{5.97} & 97.21 & \textbf{84.89} & \textbf{30.77} & 61.22 \\
    \bottomrule
    \end{tabular}
    }
\vspace{-4.5pt}
\end{table*}

\subsection{Detailed SUPERB Benchmark Results}
\label{sec:main-results}

In Table\,\ref{tab:full-superb-benchmark}, STaRHuBERT outperforms other methods\,\cite{jang23_interspeech, peng23c_interspeech} by a large margin for both 100h and 960h distillations, presenting the superiority of our distillation objectives.
Among the models with less than 24 million parameters, STaRHuBERT shows the best performance in both overall\,\cite{chen2021wavlm} and generalizability\,\cite{feng2023superb} scores with the fewest parameters.
STaRHuBERT-L, which has the wider FFN width of 1392, attains an overall score of 79.8, even surpassing LigthHuBERT\,\cite{wang2022lighthubert} for the first time as an approach that does not demand excessive computational resources for compression.

Delving into specific tasks, STaRHuBERT excels in 5 out of 10 downstream tasks compared to ARMHuBERT-S\,\cite{jang23_interspeech} and DPHuBERT\,\cite{peng23c_interspeech}.
Especially for both PR and ASR, the performance of STaRHuBERT notably surpasses these approaches, verifying that the temporal relation is crucial in learning speech representations associated with acoustic units for lightweight SSL model.
Our student models also achieve the outstanding performance in the speaker-related tasks, indicating that the TGMs can also capture speaker information by distinguishing the temporal relations of different speakers.
Taken together, our proposed STaR distillation allows us to construct a more compact model with strong performance and lightweight characteristics.

Our methodology also involves fewer MACs for inference compared to approaches that conduct both pruning and distillation.
STaRHuBERT exhibits a parameter reduction of only 5.43\% compared to DPHuBERT\,\cite{peng23c_interspeech}, but requires 29.1\% fewer MACs.
The relatively high number of MACs observed in pruning approach is also evident in the comparison between STaRHuBERT-L and LightHuBERT\,\cite{wang2022lighthubert}, wherein the former requires 40.2\% fewer MACs.
This implies that pruning technique can achieve good model performance and desired number of parameters by adjusting the sparsity ratio, however, it is challenging to control the computational cost of the pruned model.
Hence, our approach, which is one of the vanilla distillation approaches with a predefined model configuration, has a benefit in terms of efficient and lightweight inference.

\subsection{Examination on Universality}

We confirm the universality of the STaR loss by replacing the teacher model with wav2vec\,2.0\,\textsc{Base}\,\cite{baevski2020wav2vec} or wavLM\,\textsc{Base}\,\cite{chen2021wavlm}.
Table\,\ref{tab:other-teacher-superb-results-table} depicts the SUPERB benchmark results for LibriSpeech\,\cite{panayotov2015librispeech} 960h distillation.
STaRW2V2 offers stronger representation overall with only 70.5\% parameters of FitW2V2\,\cite{lee2022fithubert}.
STaRwavLM also outperforms previous methods\,\cite{jang23_interspeech, wang23da_interspeech}, even with the fewest parameters.
These student models corroborate that the proposed STaR distillation is agnostic to other Transformer-based speech SSL models.
\subsection{Compression for Smaller Parameter Sizes}

\begin{figure}[!t]
  \centering
  \vspace{3pt}
  \begin{subfigure}[b]{0.157\textwidth}
  \centering \hspace{15pt}(a) ASR
    \includegraphics[width=\linewidth]{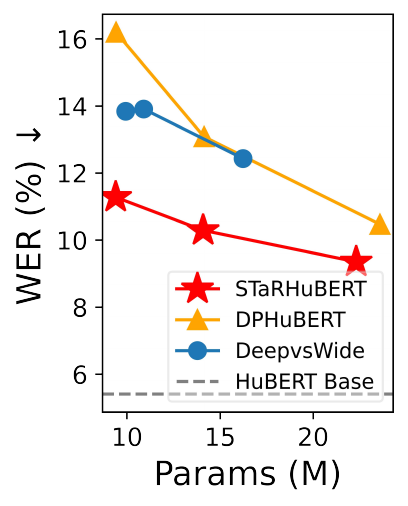}
  \end{subfigure}
  \begin{subfigure}[b]{0.157\textwidth}
  \centering \hspace{15pt}(b) SID
    \includegraphics[width=\linewidth]{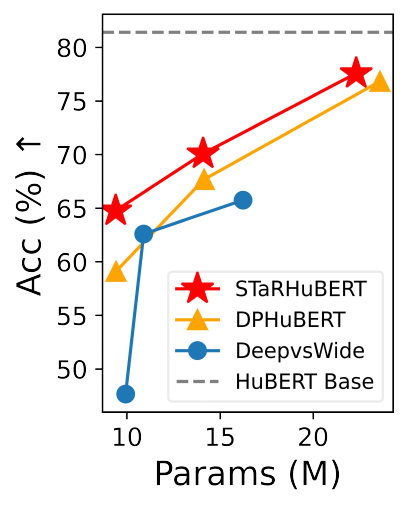}
  \end{subfigure}
  \begin{subfigure}[b]{0.157\textwidth}
  \centering \hspace{15pt}(c) IC
    \includegraphics[width=\linewidth]{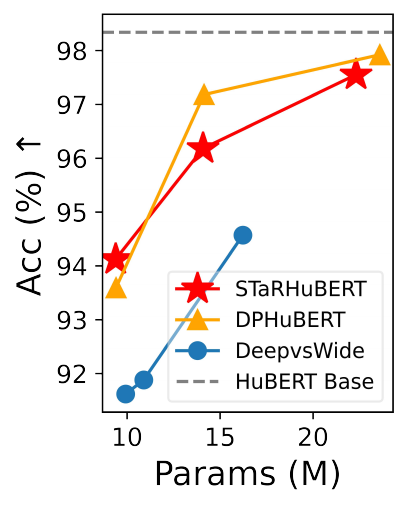}
  \end{subfigure}
  \vspace{-18pt}
  \caption{Performance comparisons of models with fewer parameters.}
  \label{fig:comparisons-of-fewer-parameters}
  \vspace{-0.25pt}
\end{figure}

To further support that our proposed STaR distillation is effective for lightweight models, we compare the performance of the models compressed to less than 20 million parameters\,\cite{peng23c_interspeech, ashihara2022deep}.
Fig.\,\ref{fig:comparisons-of-fewer-parameters} compares ASR, SID, and IC performances on the SUPERB benchmark using LibriSpeech\,\cite{panayotov2015librispeech} 960h distillation.
Our model performs best in all three tasks among the models with 9 million parameters and also shows the least performance degradation as the number of parameters decreased.
Based on our findings, distillation of speech temporal relation is preferable than directly matching the complex teacher representation for lightweight student model with limited capacity.

\begin{table}[!t]
    \centering
    \vspace{-6pt} 
    \caption{
    Comparisons of the SUPERB benchmark results for different speech SSL models, wav2vec 2.0 \textsc{Base} and wavLM \textsc{Base}.
    }
    \label{tab:other-teacher-superb-results-table}
    \vspace{4.5pt}
    \addtolength{\tabcolsep}{2pt}
    \resizebox{\linewidth}{!}{
    \begin{tabular}{lccc}
    \toprule
    Models & Params\,(M)\,$\downarrow$ & Overall\,$\uparrow$ & General.\,$\uparrow$ \\
    \midrule
    wav2vec 2.0\,\textsc{Base}\,\cite{baevski2020wav2vec} & 95.04 & 79.0 & 818 \\ 
    \midrule
    FitW2V2\,\cite{lee2022fithubert}  & 31.63 & 76.5 & 766 \\
    \textbf{STaRW2V2\,(ours)}  & \textbf{22.31} & \textbf{77.2} & \textbf{797} \\
    \midrule[0.8pt]
    wavLM\,\textsc{Base}\,\cite{chen2021wavlm} & 94.70 & 81.9 & 1019 \\
    \midrule
    Structured Pruning\,\cite{wang23da_interspeech} & 26.57 & 78.9 & 863 \\
    ARMwavLM-S\,\cite{jang23_interspeech}  & 22.39 & 78.9 & 851 \\
    \textbf{STaRwavLM\,(ours)}  & \textbf{22.31} & \textbf{79.4} & \textbf{899} \\
    \bottomrule
    \end{tabular}
    }
\vspace{-4.5pt}
\end{table}

\section{Conclusion}

In this study, we have proposed to distill speech temporal relation, STaR, from the Transformer-based speech SSL models.
We combine the distillation of layer-wise and intra-layer TGMs as the STaR loss, which does not require any additional parameters for distillation.
STaRHuBERT-L achieves the SOTA overall score of 79.8 among models with $\sim$\,27 million parameters, in particular showing favorable results on PR, ASR, and speaker-related tasks.
Further experiments reveal that our method is agnostic to other speech SSL models and is more suitable to lightweight students.
To sum up, our results present the effectiveness of distilling STaR rather than directly matching the output representation for lightweight SSL models.

\vfill
\clearpage
\newpage
\pagebreak

\section{References}
\bibliographystyle{IEEEbib}
{\footnotesize
    \bibliography{strings,refs}}

\begin{thebibliography}{10}

\bibitem{baevski2020wav2vec}
Alexei Baevski, Yuhao Zhou, Abdelrahman Mohamed, and Michael Auli,
\newblock ``wav2vec 2.0: A framework for self-supervised learning of speech representations,''
\newblock in {\em Proc. NeurIPS}, 2020, pp. 12449--12460.

\bibitem{hsu2021hubert}
Wei-Ning Hsu, Benjamin Bolte, Yao-Hung~Hubert Tsai, Kushal Lakhotia, Ruslan Salakhutdinov, and Abdelrahman Mohamed,
\newblock ``Hubert: Self-supervised speech representation learning by masked prediction of hidden units,''
\newblock {\em IEEE/ACM Transactions on Audio, Speech, and Language Processing}, vol. 29, pp. 3451--3460, 2021.

\bibitem{chen2021wavlm}
Sanyuan Chen, Chengyi Wang, Zhengyang Chen, Yu~Wu, Shujie Liu, Zhuo Chen, Jinyu Li, Naoyuki Kanda, Takuya Yoshioka, Xiong Xiao, et~al.,
\newblock ``Wavlm: Large-scale self-supervised pre-training for full stack speech processing,''
\newblock {\em IEEE Journal of Selected Topics in Signal Processing}, vol. 16, no. 6, pp. 1505--1518, 2022.

\bibitem{yang21c_interspeech}
Shu wen Yang, Po-Han Chi, Yung-Sung Chuang, Cheng-I~Jeff Lai, Kushal Lakhotia, Yist~Y. Lin, Andy~T. Liu, Jiatong Shi, Xuankai Chang, Guan-Ting Lin, Tzu-Hsien Huang, Wei-Cheng Tseng, Ko~tik Lee, Da-Rong Liu, Zili Huang, Shuyan Dong, Shang-Wen Li, Shinji Watanabe, Abdelrahman Mohamed, and Hung yi~Lee,
\newblock ``Superb: Speech processing universal performance benchmark,''
\newblock in {\em Proc. Interspeech}, 2021, pp. 1194--1198.

\bibitem{feng2023superb}
Tzu-hsun Feng, Annie Dong, Ching-Feng Yeh, Shu-wen Yang, Tzu-Quan Lin, Jiatong Shi, Kai-Wei Chang, Zili Huang, Haibin Wu, Xuankai Chang, et~al.,
\newblock ``Superb@ slt 2022: Challenge on generalization and efficiency of self-supervised speech representation learning,''
\newblock in {\em Proc. SLT}, 2023, pp. 1096--1103.

\bibitem{lai2021parp}
Cheng-I~Jeff Lai, Yang Zhang, Alexander~H Liu, Shiyu Chang, Yi-Lun Liao, Yung-Sung Chuang, Kaizhi Qian, Sameer Khurana, David Cox, and Jim Glass,
\newblock ``Parp: Prune, adjust and re-prune for self-supervised speech recognition,''
\newblock in {\em Proc. NeurIPS}, 2021, pp. 21256--21272.

\bibitem{wang2022deep}
Naigang Wang, Chi-Chun~(Charlie) Liu, Swagath Venkataramani, Sanchari Sen, Chia-Yu Chen, Kaoutar El~Maghraoui, Vijayalakshmi~(Viji) Srinivasan, and Leland Chang,
\newblock ``Deep compression of pre-trained transformer models,''
\newblock in {\em Proc. NeurIPS}, 2022, pp. 14140--14154.

\bibitem{choi2023masked}
Euntae Choi, Youshin Lim, Byeong-Yeol Kim, Hyung~Yong Kim, Hanbin Lee, Yunkyu Lim, Seung~Woo Yu, and Sungjoo Yoo,
\newblock ``Masked token similarity transfer for compressing transformer-based asr models,''
\newblock in {\em Proc. ICASSP}, 2023, pp. 1--5.

\bibitem{lim2023lightweight}
Hyungjun Lim, Younggwan Kim, Kiho Yeom, Eunjoo Seo, Hoodong Lee, Stanley~Jungkyu Choi, and Honglak Lee,
\newblock ``Lightweight feature encoder for wake-up word detection based on self-supervised speech representation,''
\newblock in {\em Proc. ICASSP}, 2023, pp. 1--5.

\bibitem{heo23_interspeech}
Jungwoo Heo, Chan yeong Lim, Ju~ho~Kim, Hyun seo Shin, and Ha-Jin Yu,
\newblock ``{One-Step Knowledge Distillation and Fine-Tuning in Using Large Pre-Trained Self-Supervised Learning Models for Speaker Verification},''
\newblock in {\em Proc. Interspeech}, 2023, pp. 5271--5275.

\bibitem{chang2021distilhubert}
Heng-Jui Chang, Shu-wen Yang, and Hung-yi Lee,
\newblock ``Distilhubert: Speech representation learning by layer-wise distillation of hidden-unit bert,''
\newblock in {\em Proc. ICASSP}, 2022, pp. 7087--7091.

\bibitem{lee2022fithubert}
Yeonghyeon Lee, Kangwook Jang, Jahyun Goo, Youngmoon Jung, and Hoirin Kim,
\newblock ``Fithubert: Going thinner and deeper for knowledge distillation of speech self-supervised learning,''
\newblock in {\em Proc. Interspeech}, 2022, pp. 3588--3592.

\bibitem{jang23_interspeech}
Kangwook Jang, Sungnyun Kim, Se-Young Yun, and Hoirin Kim,
\newblock ``{Recycle-and-Distill: Universal Compression Strategy for Transformer-based Speech SSL Models with Attention Map Reusing and Masking Distillation},''
\newblock in {\em Proc. Interspeech}, 2023, pp. 316--320.

\bibitem{vaswani2017attention}
Ashish Vaswani, Noam Shazeer, Niki Parmar, Jakob Uszkoreit, Llion Jones, Aidan~N Gomez, \L~ukasz Kaiser, and Illia Polosukhin,
\newblock ``Attention is all you need,''
\newblock in {\em Proc. NeurIPS}, 2017, pp. 6000--6010.

\bibitem{peng23c_interspeech}
Yifan Peng, Yui Sudo, Shakeel Muhammad, and Shinji Watanabe,
\newblock ``Dphubert: Joint distillation and pruning of self-supervised speech models,''
\newblock in {\em Proc. Interspeech}, 2023, pp. 62--66.

\bibitem{wang23da_interspeech}
Haoyu Wang, Siyuan Wang, Wei-Qiang Zhang, Suo Hongbin, and Yulong Wan,
\newblock ``Task-agnostic structured pruning of speech representation models,''
\newblock in {\em Proc. Interspeech}, 2023, pp. 231--235.

\bibitem{louizos2018learning}
Christos Louizos, Max Welling, and Diederik~P Kingma,
\newblock ``Learning sparse neural networks through l\_0 regularization,''
\newblock in {\em Proc. ICLR}, 2018.

\bibitem{wang2022lighthubert}
Rui Wang, Qibing Bai, Junyi Ao, Long Zhou, Zhixiang Xiong, Zhihua Wei, Yu~Zhang, Tom Ko, and Haizhou Li,
\newblock ``Lighthubert: Lightweight and configurable speech representation learning with once-for-all hidden-unit bert,''
\newblock in {\em Proc. Interspeech}, 2022.

\bibitem{ashihara2022deep}
Takanori Ashihara, Takafumi Moriya, Kohei Matsuura, and Tomohiro Tanaka,
\newblock ``Deep versus wide: An analysis of student architectures for task-agnostic knowledge distillation of self-supervised speech models,''
\newblock in {\em Proc. Interspeech}, 2022, pp. 411--415.

\bibitem{baevski2021unsupervised}
Alexei Baevski, Wei-Ning Hsu, Alexis Conneau, and Michael Auli,
\newblock ``Unsupervised speech recognition,''
\newblock in {\em Proc. NeurIPS}, 2021, pp. 27826--27839.

\bibitem{wang2020minilm}
Wenhui Wang, Furu Wei, Li~Dong, Hangbo Bao, Nan Yang, and Ming Zhou,
\newblock ``Minilm: Deep self-attention distillation for task-agnostic compression of pre-trained transformers,''
\newblock in {\em Proc. NeurIPS}, 2020, pp. 5776--5788.

\bibitem{fang2021compressing}
Zhiyuan Fang, Jianfeng Wang, Xiaowei Hu, Lijuan Wang, Yezhou Yang, and Zicheng Liu,
\newblock ``Compressing visual-linguistic model via knowledge distillation,''
\newblock in {\em Proc. ICCV}, 2021, pp. 1428--1438.

\bibitem{gatys2015texture}
Leon Gatys, Alexander~S Ecker, and Matthias Bethge,
\newblock ``Texture synthesis using convolutional neural networks,''
\newblock in {\em Proc. NeurIPS}, 2015, pp. 262--270.

\bibitem{stafylakis21_interspeech}
Themos Stafylakis, Johan Rohdin, and Lukáš Burget,
\newblock ``Speaker embeddings by modeling channel-wise correlations,''
\newblock in {\em Proc. Interspeech}, 2021, pp. 501--505.

\bibitem{saito2021perceptual}
Yuki Saito, Shinnosuke Takamichi, and Hiroshi Saruwatari,
\newblock ``Perceptual-similarity-aware deep speaker representation learning for multi-speaker generative modeling,''
\newblock {\em IEEE/ACM Transactions on Audio, Speech, and Language Processing}, vol. 29, pp. 1033--1048, 2021.

\bibitem{li2020non}
Yanping Li, Dongxiang Xu, Yan Zhang, Yang Wang, and Binbin Chen,
\newblock ``Non-parallel many-to-many voice conversion with psr-stargan.,''
\newblock in {\em Proc. Interspeech}, 2020, pp. 781--785.

\bibitem{shi2022u}
Sheng Shi, Jiahao Shao, Yifei Hao, Yangzhou Du, and Jianping Fan,
\newblock ``U-gat-vc: Unsupervised generative attentional networks for non-parallel voice conversion,''
\newblock in {\em Proc. ICASSP}, 2022, pp. 7017--7021.

\bibitem{gretton2012kernel}
Arthur Gretton, Karsten~M Borgwardt, Malte~J Rasch, Bernhard Sch{\"o}lkopf, and Alexander Smola,
\newblock ``A kernel two-sample test,''
\newblock {\em The Journal of Machine Learning Research}, vol. 13, pp. 723--773, 2012.

\bibitem{li2017demystifying}
Yanghao Li, Naiyan Wang, Jiaying Liu, and Xiaodi Hou,
\newblock ``Demystifying neural style transfer,''
\newblock in {\em Proc. IJCAI}, 2017, pp. 2230--2236.

\bibitem{yim2017gift}
Junho Yim, Donggyu Joo, Jihoon Bae, and Junmo Kim,
\newblock ``A gift from knowledge distillation: Fast optimization, network minimization and transfer learning,''
\newblock in {\em Proc. CVPR}, 2017, pp. 4133--4141.

\bibitem{zhou2022efficient}
ZhaoJing Zhou, Yun Zhou, Zhuqing Jiang, Aidong Men, and Haiying Wang,
\newblock ``An efficient method for model pruning using knowledge distillation with few samples,''
\newblock in {\em Proc. ICASSP}, 2022, pp. 2515--2519.

\bibitem{peng2019correlation}
Baoyun Peng, Xiao Jin, Jiaheng Liu, Dongsheng Li, Yichao Wu, Yu~Liu, Shunfeng Zhou, and Zhaoning Zhang,
\newblock ``Correlation congruence for knowledge distillation,''
\newblock in {\em Proc. ICCV}, 2019, pp. 5007--5016.

\bibitem{jin2023cross}
Yufeng Jin, Guosheng Hu, Haonan Chen, Duoqian Miao, Liang Hu, and Cairong Zhao,
\newblock ``Cross-modal distillation for speaker recognition,''
\newblock in {\em Proc. AAAI}, 2023, pp. 12977--12985.

\bibitem{pasad2023comparative}
Ankita Pasad, Bowen Shi, and Karen Livescu,
\newblock ``Comparative layer-wise analysis of self-supervised speech models,''
\newblock in {\em Proc. ICASSP}, 2023, pp. 1--5.

\bibitem{panayotov2015librispeech}
Vassil Panayotov, Guoguo Chen, Daniel Povey, and Sanjeev Khudanpur,
\newblock ``Librispeech: an asr corpus based on public domain audio books,''
\newblock in {\em Proc. ICASSP}, 2015, pp. 5206--5210.

\end{thebibliography}
\end{document}